# Selective stabilization of antiferromagnetic orders in FeTe films via local strain engineering

Hao Xu[1,4,5,*], Jing Jiang[2,3,*], Xuesong Gai[1,4,5], Haicheng Lin[1], Kai Liu[2,3]✉, Zhong-Yi Lu[2,3], Kai Chang[1]✉ & Chong Liu[1]✉

*[1] Beijing Key Laboratory of Fault-Tolerant Quantum Computing, Beijing Academy of Quantum Information Sciences, Beijing 100193, China*

*[2] School of Physics, Renmin University of China, Beijing 100872, China*

*[3] Key Laboratory of Quantum State Construction and Manipulation (Ministry of Education), Renmin University of China, Beijing 100872, China*

*[4] Beijing National Laboratory for Condensed Matter Physics, Institute of Physics, Chinese Academy of Sciences, Beijing 100190, China*

*[5] University of Chinese Academy of Sciences, Beijing 100049, China*

*These authors contributed equally

✉e-mail: kliu@ruc.edu.cn; changkai@baqis.ac.cn; liuchong@baqis.ac.cn

## Abstract

The parent compound FeTe hosts a complex magnetic landscape that is highly susceptible to lattice distortions. Although theoretical models have predicted a bicollinear to dimer antiferromagnetic (AFM) phase transition under tensile strain, its experimental realization and deterministic control has remained elusive owing to severe magnetic frustration. Here, combining high-resolution scanning tunneling microscopy (STM) and density functional theory (DFT) calculations, we demonstrate the selective stabilization of bicollinear and dimer AFM orders in few-layer FeTe films via local uniaxial strain engineering. By mapping the strain fields near dislocation areas in FeTe films and FeTe/FeSe heterostructures, we establish a direct correspondence between specific strain components and the resulting magnetic ground states. We find that uniaxial compression along the Fe-Fe next-nearest-neighbor direction stabilizes the bicollinear AFM order, with the stripe orientation aligning parallel to the compression axis. Crucially, we report the experimental realization of the long-range dimer AFM order, which emerges under anisotropic strain along the Fe-Fe nearest-neighbor direction. This phase manifests as a distinct $\sqrt{2} \times \sqrt{2}$ electronic reconstruction and shares a common Néel temperature with the bicollinear phase. Our findings reveal that anisotropic strain effectively lifts the magnetic degeneracy among competing states. This work provides a robust strategy for the manipulation of elusive magnetic orders and offers insights into the interplay between lattice, spin, and electronic degrees of freedom in iron-based superconductors.

## Introduction

The search for unconventional superconductivity and exotic quantum phases has centered on the intricate interplay between lattice, spin, and orbital degrees of freedom. In the family of iron-based superconductors (IBSs), the parent compound FeTe stands out as a unique platform due to its enigmatic (π/2, π/2) bicollinear antiferromagnetic (AFM) ground state, which contrasts sharply with the (π, 0) stripe order found in iron pnictides. This magnetic diversity is rooted in a highly frustrated energy landscape where multiple magnetic configurations, such as bicollinear, dimer, and trimer states, reside in close energetic proximity [1]. Consequently, FeTe exhibits an extraordinary sensitivity to external tuning parameters, particularly lattice strain, which can tip the balance between these nearly degenerate states [2-4].

Strain engineering has emerged as a powerful, non-invasive laboratory to explore hidden phases and manipulate electronic correlation [5-7]. While hydrostatic pressure has been extensively studied, it often preserves the crystal's rotational symmetry, potentially masking phases that emerge only upon symmetry breaking [5]. In contrast, uniaxial strain provides a surgical tool to lift the $C_4$ rotational symmetry, directly addressing the spin-nematicity and orbital-selective phenomena characteristic of iron chalcogenides [8]. Despite theoretical predictions of strain-induced "n-mer" magnetic states in FeSe and FeTe [9-11], capturing these phases in real space and establishing a deterministic correspondence with specific strain components remains a formidable challenge.

In this work, we focus on ultrathin FeTe films that suffer strong strain fluctuation. We utilize the localized strain fields surrounding inherent edge dislocations and the interface induced dislocation lines to systematically map the magnetic phase of FeTe. By combining atomic-resolution scanning tunneling microscopy (STM) with geometric phase analysis (GPA) and density functional theory (DFT) calculations, we provide direct evidence of how directional strain selectively stabilizes either bicollinear or dimer AFM orders, offering a microscopic view of how symmetry-breaking strain resolves magnetic frustration.

## Results

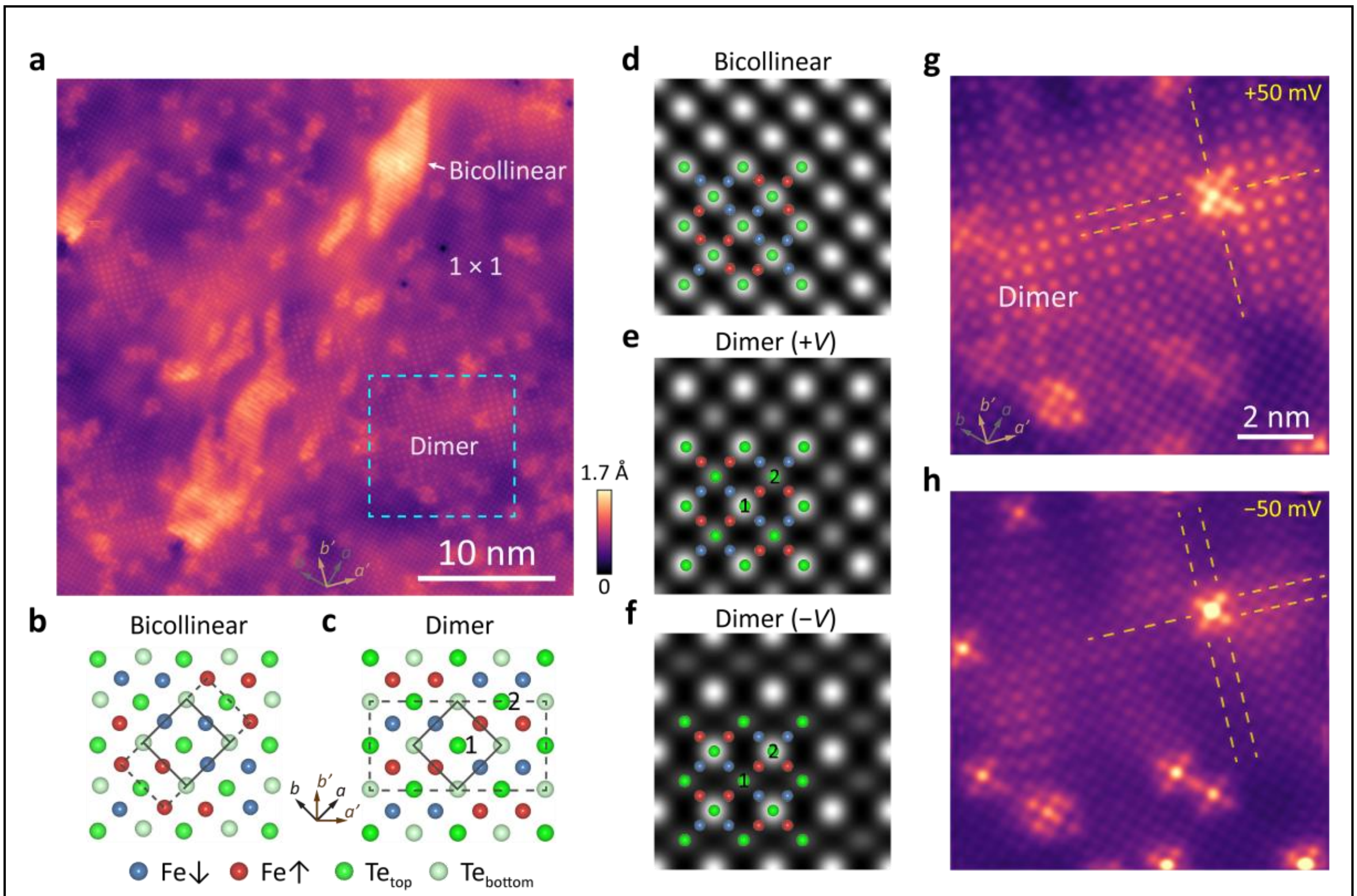

**Figure 1. Real-space observation of bicollinear and dimer AFM orders in bilayer FeTe.** (a) Typical STM topographic image of bilayer FeTe/STO taken at 4.6 K, revealing local stripe-like and $\sqrt{2} \times \sqrt{2}$ modulated regions coexisting on the 1 × 1 lattice background. (b, c) Schematic of the bicollinear AFM (BiAFM) order and the dimer AFM order, respectively. The solid boxes indicate the structural unit cells while the dashed boxes indicate the magnetic unit cells. (d) Simulated STM image of the BiAFM phase. (e, f) Simulated STM images of the dimer AFM phase at positive (+300 mV) and negative (−300 mV) bias voltages, respectively, showing a $\sqrt{2} \times \sqrt{2}$ order with opposite intensity contrast on the two Te subsets. Bottom Te are omitted in the overlaying lattice model. (g, h) Experimental bias-dependent STM images of the $\sqrt{2} \times \sqrt{2}$ region corresponding to the square area in (a). The dashed lines manifest the higher-intensity Te subsets around an interstitial Fe impurity.

## Probing local magnetic orders via STM and DFT simulations

To investigate the magnetic ground state under strain, we performed high-resolution STM measurements on bilayer FeTe films on STO. The films underwent Te-vapor annealing to minimize interstitial Fe impurities, thereby allowing the magnetically frustrated state to dominate[11,12]. Different from the thicker FeTe films, however, few-layer FeTe films host pronounced microscopic strain variations that facilitate the emergence of nanoscale AFM orders. As shown in **Fig. 1a**, the topographic image clearly exhibits two distinct types of local reconstructions embedded in the 1 × 1 Te-terminated lattice: a stripe-like phase and a $\sqrt{2} \times \sqrt{2}$ modulated phase.

The stripe-like feature corresponds to the well-known bicollinear AFM (BiAFM) order[11]. In this configuration, Fe spins form ferromagnetic chains along one of the Fe-Fe next nearest neighboring (NNN) directions, while alternating antiferromagnetically along the orthogonal axis (**Fig. 1b**). The simulated STM image for the BiAFM phase (**Fig. 1d**) shows excellent agreement with experimental stripe patterns, where the charge density of surface Te atoms is broadened along the *b* direction.

More importantly, we focus on the $\sqrt{2} \times \sqrt{2}$ regions, which are identified as the dimer AFM order hitherto unseen in FeTe. This phase consists of Fe-Fe nearest-neighbor (NN) pairs that are ferromagnetically coupled internally but antiferromagnetically coupled to neighboring dimers (**Fig. 1c**). This specific spin arrangement breaks the structural symmetry and generates a periodic electronic reconstruction. Our bias-dependent STM images (**Fig. 1g-h**) reveal that the surface Te atoms in this region split into two sublattices. Remarkably, the relative tunneling intensity of these two Te groups undergoes a contrast inversion upon reversing the bias voltage. The reconstruction gets vanished with increasing bias voltage (**Fig. S1**), confirming its electronic origin. This $\sqrt{2} \times \sqrt{2}$ electronic ordering and its contrast flip are perfectly captured by our DFT-based STM simulations (**Fig. 1e-f**), providing conclusive evidence for the dimer AFM order. This constitutes the direct experimental observation of dimer AFM in the FeTe system, which was theoretically predicted to be a metastable ground state under tensile strain[11] but is stabilized here within localized areas of the ultrathin film.

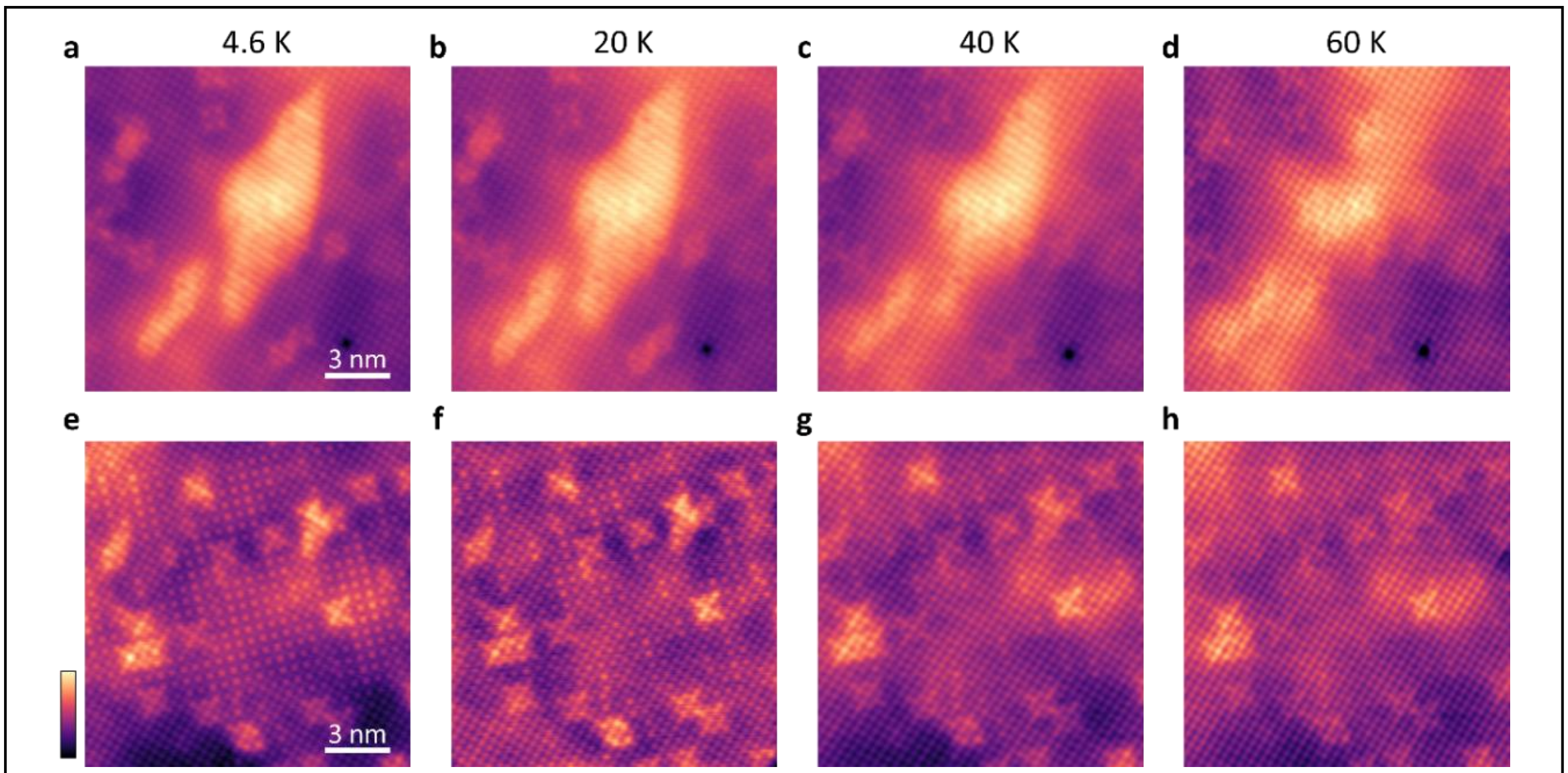

**Figure 2. Temperature-dependent evolution of two AFM orders.** (a–d) Temperature-dependent STM topographic images of the BiAFM region, showing the gradual suppression of the stripe-like modulation with increasing temperature. (e–g) Corresponding temperature-dependent STM images of the dimer AFM region. Both the bicollinear stripe and the $\sqrt{2} \times \sqrt{2}$ modulations vanish simultaneously around $T \approx 60$ K.

To further substantiate the magnetic nature of the observed $\sqrt{2} \times \sqrt{2}$ phase, we investigated the thermal evolution of both the BiAFM and dimer AFM superstructures. **Figures 2a–d** display the STM images of the BiAFM region at various temperatures. The characteristic bicollinear stripes weaken significantly as the temperature rises, eventually fading into the 1 × 1 lattice background at around 60 K. This temperature corresponds well to the Néel temperature ($T_N$) of bulk or thick FeTe films[13,14].

Crucially, the $\sqrt{2} \times \sqrt{2}$ modulation in the dimer AFM region exhibits a nearly identical temperature dependence (**Fig. 2e–h**). The electronic signature of the dimer AFM diminishes in the same temperature range as the BiAFM, vanishing completely around 60 K. The synchronized thermal melting of the two distinct spatial modulations indicates that the $\sqrt{2} \times \sqrt{2}$ order is indeed a long-range antiferromagnetic state. The fact that the dimer AFM phase shares the same $T_N$ as the well-established BiAFM phase strongly suggests that they are competing magnetic ground states of the same underlying spin system, locally stabilized by subtle variations in environmental conditions.

## Strain selectivity of the AFM orderings

To disentangle various magnetic ground states, we utilized the strain field surrounding an edge dislocation as a nanoscopic probe. As illustrated in **Fig. 3a**, an edge dislocation introduces pronounced lattice distortions, involving regions of compressive, tensile, and shear strain. High-resolution STM imaging in the vicinity of a dislocation core (**Fig. 3b**) reveals a striking spatial segregation of magnetic phases: a BiAFM region and dimer AFM regions coexist in close proximity but reside on different sides of the dislocation core.

We quantified the local lattice strain tensor via the geometric phase analysis (GPA) method. Here, the $a$ and $b$ axes are defined along Fe-Fe NNN directions, whereas the $a'$ and $b'$ axes are aligned with the Fe-Fe NN directions. The extra half column of atoms inserted on the upper side of the dislocation core induces a strong compressive (tensile) strain along the $a$ direction on the upper (lower) side of the core (**Fig. 3c**), but having a negligible effect on the strain along the $b$ direction (**Fig. 3d**), as expected. By comparing with the STM topography in **Fig. 3b**, we observe that the BiAFM phase predominantly emerges in the zone experiencing uniaxial compression along the $a$ axis (blue dashed lines).

In contrast, the dimer AFM phase develops in areas characterized by a substantial shear strain $\varepsilon_{ab}$ in the *ab* coordinate system (**Fig. 3b, 3e, 3f**), flanking the left and right sides of the core (green dashed lines). The $2\varepsilon_{ab}$ is mathematically equivalent to the anisotropic normal strain along the $a'$-$b'$ axes, expressed as $\varepsilon_{b'b'} - \varepsilon_{a'a'}$ (see Supporting Materials Note I and **Fig. S3** for the derivation and demonstration). Therefore, the formation of the dimer AFM is correlated with the unequal lattice distortion between the two orthogonal Fe-Fe NN directions.

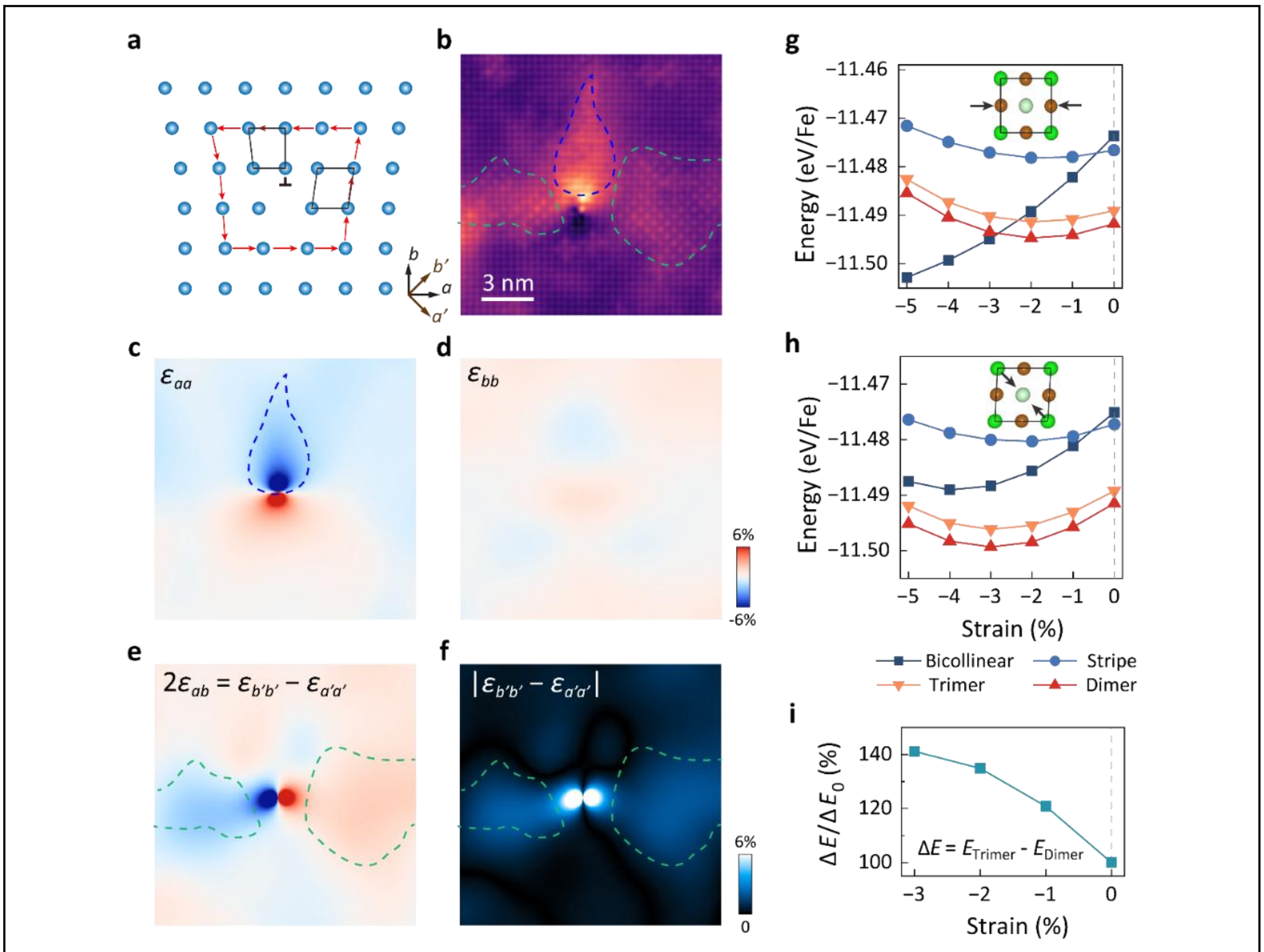

**Figure 3. Strain-induced selectivity of AFM orders near an edge dislocation.** (a) Schematic of an edge dislocation in a 2D lattice. The blue balls denote the top-layer Te atoms. The red arrows represent the Burgers circuit, and the black polygons illustrate the locally distorted unit cells. (b) Atomic-resolution STM image of the region surrounding an edge dislocation. Dashed loops highlight the spatially separated BiAFM and dimer AFM regions. (c–e) Strain tensor distribution maps for (b) obtained via geometric phase analysis (GPA), showing different strain components ($\varepsilon_{aa}$, $\varepsilon_{bb}$ and $2\varepsilon_{ab}$). (f) The absolute value of (e). (g–i) DFT calculated energies of various magnetic states as a function of strain. (g) Energy versus uniaxial compression along the *a* or *b* axis (Fe-Fe NNN direction). The BiAFM state becomes the ground state when compression exceeds 3%. (h) Energy versus uniaxial compression along the $a'$ or $b'$ axis (Fe-Fe NN direction). The dimer AFM state remains the lowest in energy. (i) Relative energy difference between the dimer and trimer AFM states in (h), showing that strain significantly widens the energy gap.

These experimental findings are fully corroborated by our DFT calculations. We initiate our analysis by considering uniformly in-plane tensile strained FeTe with the lattice constant of 3.9 Å, in which the dimer and trimer AFM states have lower energy than the BiAFM state[11]. When uniaxial compression is applied along the *a* or *b* axis, the total energy of the BiAFM state drops rapidly, establishing it as the stable configuration at beyond ~3% strain (**Fig. 3g**). Conversely, compression along the $a'$ or $b'$ axis preserves the dimer AFM as the ground state (**Fig. 3h**). Notably, while our previous work indicated that the dimer and trimer states are nearly degenerate under isotropic tensile strain[11], we demonstrate here that a ~3% uniaxial strain along the Fe-Fe NN direction effectively widens the energy splitting between these two states (**Fig. 3i**). This strain-induced lifting of degeneracy suppresses magnetic frustration, thereby facilitating the long-range $\sqrt{2} \times \sqrt{2}$ dimer AFM order observed by STM.

To further establish the deterministic control and universality of strain-induced magnetic regulation, we fabricated a 4-ML FeTe/2-ML FeSe/STO heterostructure. In this architecture, the robust dislocation lines inherently present in the 2-ML FeSe buffer layer[15] act as a template to propagate local strain into the overlying FeTe film. As a result, the morphology on the top layer FeTe (**Fig. 4a**) displays strong dislocation grids, highly reminiscent of those observed in pristine FeSe films[16]. This interface enhanced strain field provides an ideal platform to unambiguously resolve how distinct strain components dictate the type of magnetic ordering.

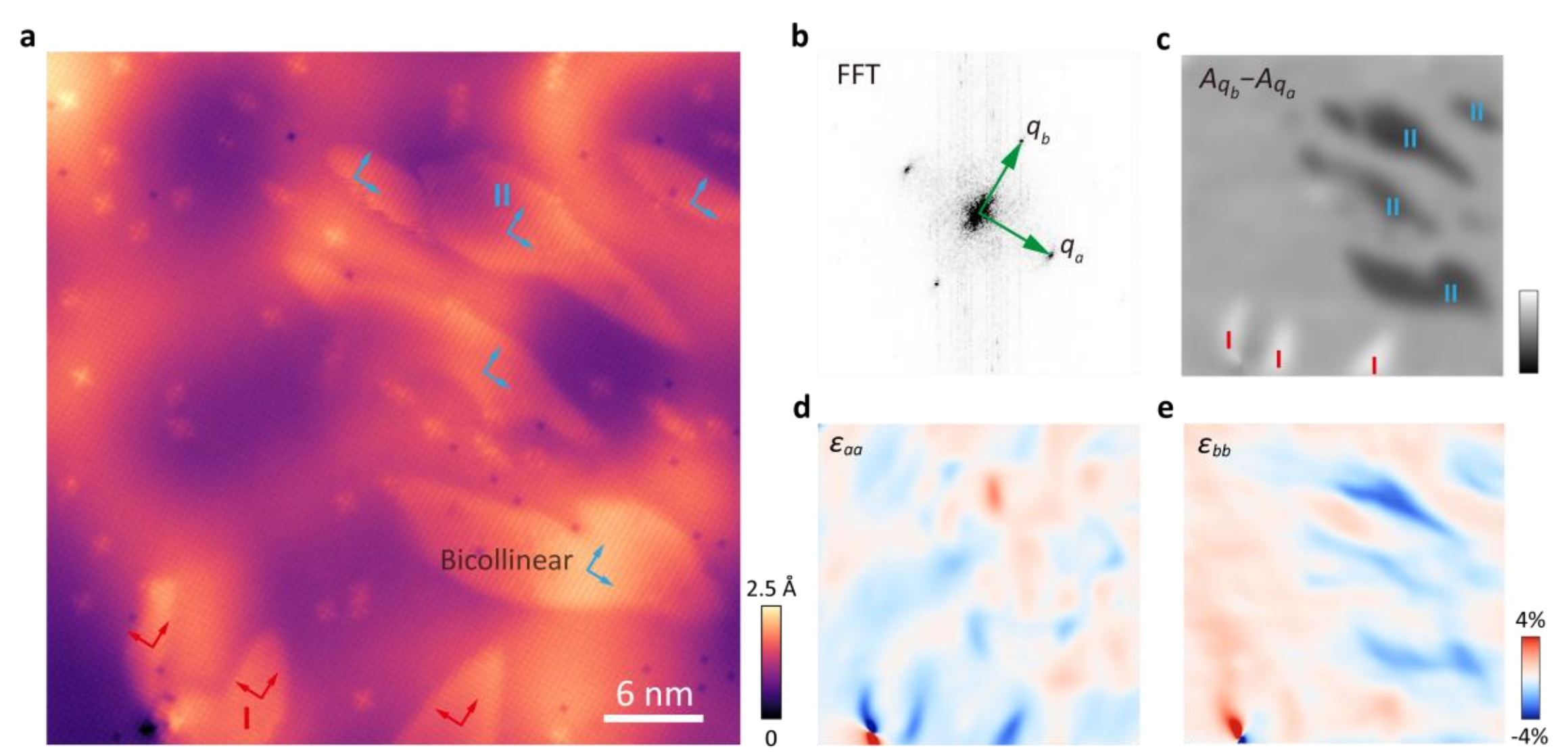

**Figure 4. Deterministic control of bicollinear AFM domains via uniaxial strain.** (a) Atomic-resolution STM topographic image of 4-ML FeTe/2-ML FeSe/STO. Regions I and II represent two distinct bicollinear AFM domains with orthogonal stripe orientations. (b) Fourier transform (FT) of the image in (a), where $q_a$ and $q_b$ denote the reciprocal lattice vectors. (c) Spatial map showing the difference in local intensity between $q_a$ and $q_b$, obtained using a 2D lock-in technique. The contrast clearly distinguishes the spatial distribution of the two domain types. (d, e) Corresponding local strain maps ($\varepsilon_{aa}$ and $\varepsilon_{bb}$) extracted from (a) via GPA method.

As shown in **Fig. 4a**, the STM topography reveals two orthogonal BiAFM domains, labeled as Region I and Region II. The origin of these stripes lies in the directional broadening of electronic states along either the $a$ or $b$ axis of the 1 × 1 lattice background. This anisotropic broadening in real space leads to a selective suppression of the corresponding Bragg peak intensity in reciprocal space. By employing a two-dimensional lock-in technique to map the local amplitude difference between the $q_a$ and $q_b$ peaks ($A_{q_b} - A_{q_a}$)[11], we successfully visualized the distribution of these domains (**Fig. 4c**). In this difference map, Region I (where $q_b$ is suppressed) and Region II (where $q_a$ is suppressed) manifest as regions with values higher and lower than the isotropic 1 × 1 background, respectively, yielding a high-contrast spatial representation of the magnetic texture.

Quantitative strain analysis using GPA (**Fig. 4d,e**) uncovers a striking correlation: the stripe orientation of the BiAFM order consistently aligns with the axis of uniaxial compression. Specifically, Region I (II), featuring stripes oriented along the $a$ ($b$) direction, corresponds well to the areas with pronounced $a$-axis ($b$-axis) compression. Our DFT calculations confirm this selective stabilization, showing that uniaxial compression along a Fe-Fe NNN direction reduces the energy of the bicollinear state possessing ferromagnetic chains along the compressed axis at a much higher rate than the orthogonal orientation. For the strain between 3% and 5%, only the former configuration drops below the dimer and trimer AFM states in energy, thus becoming uniquely stabilized ("bicollinear-a" in **Fig. S7a**). These results provide conclusive evidence that the magnetic anisotropy and domain structure in FeTe can be manipulated through directional strain engineering.

We extend our investigation to the modulation of the dimer AFM order in the 2-ML FeTe/2-ML FeSe/STO heterostructure. As shown in **Fig. 5a**, a $\sqrt{2} \times \sqrt{2}$ dimer AFM phase is localized near specific dislocation-induced wrinkles. The periodicity is confirmed by the Fourier transform (**Fig. 5b**), which exhibits clear Bragg peaks corresponding to the superstructure. To visualize the spatial distribution of this phase, we performed a wavelet transform, with the blue-masked regions in **Fig. 5c** identifying the long-range dimer-ordered domains.

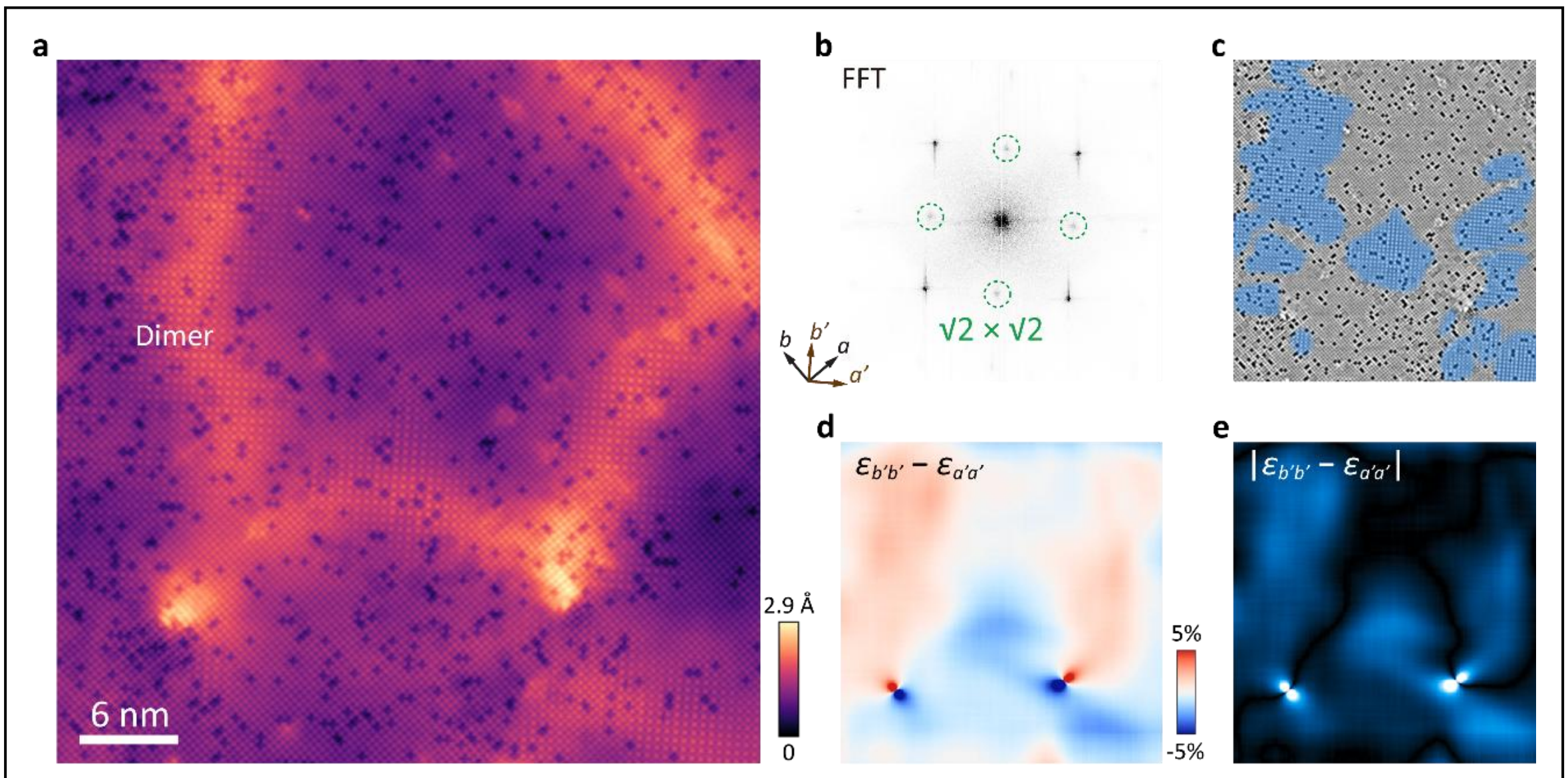

**Figure 5. Anisotropic strain regulation of the dimer AFM order in 2-ML FeTe/2-ML FeSe/STO.** (a) Atomic-resolution STM topographic image showing the emergence of the dimer AFM order in the vicinity of dislocation lines. (b) Fourier transform (FT) of (a); dashed circles indicate the $\sqrt{2} \times \sqrt{2}$ reciprocal lattice vectors. (c) Spatial distribution of the $\sqrt{2} \times \sqrt{2}$ phase, extracted via wavelet transform and highlighted with a blue mask. (d) Map of the strain anisotropy along the Fe-Fe nearest-neighbor direction, calculated as the difference between strain components $\varepsilon_{b'b'} - \varepsilon_{a'a'}$. (e) Absolute value of the strain anisotropy in (d), which shows strong correlation with (c).

The relationship between this magnetic order and the local lattice distortion was established by mapping the strain anisotropy, defined as $\varepsilon_{b'b'} - \varepsilon_{a'a'}$, along the Fe-Fe NN directions (**Fig. 5d**). The absolute value of the anisotropy map (**Fig. 5e**) reveals a striking spatial coincidence with the dimer AFM domains, echoing our observations near the edge dislocation in **Fig. 3f**. Notably, while the top-right region of **Fig. 5a** also displays a significant topographic wrinkle, the corresponding GPA analysis in **Fig. 5e** indicates minimal anisotropy falling on the $a'$-$b'$ axes. Consequently, this area remains in the $1 \times 1$ phase (**Fig. 5c**). This observation rigorously demonstrates that the formation of the dimer AFM state is uniquely coupled to the anisotropic strain along the Fe-Fe NN direction.

Additionally, our DFT calculations reveal that the dimer AFM state also lifts its degeneracy under uniaxial strain, where the ferromagnetic coupled Fe-Fe dimers tend to align parallel to the axis of uniaxial compression ("dimer-a" in **Fig. S7a**). Although this suggests the potential coexistence of two mutually orthogonal dimer domains, corresponding to the positive and negative anisotropic regions in **Fig. 5d**, they are experimentally indistinguishable in the STM measurements. This is because the top-layer Te electronic states associated with the dimer AFM order preserve a nearly four-fold symmetry in the $\sqrt{2} \times \sqrt{2}$ configuration (**Fig. 1e-f**), effectively masking the underlying rotational symmetry breaking of the Fe sublattice.

## Discussion

The deterministic stabilization of spatially segregated magnetic orders near lattice distortions provides a profound insight into the selective coupling between strain anisotropy and magnetic exchange interactions. In the isotropically strained case, theoretical models suggest that the energy landscape is nearly flat between the dimer and trimer AFM states, leading to magnetic fluctuations that suppress long-range magnetic order [11]. To understand the underlying mechanism of the emergence of dimer AFM state under uniaxial compressive strain along the Fe–Fe NN direction (**Figs. 3h,3i**), we calculated the exchange interactions among Fe spins in ML FeTe with and without strain using the $J_1$-$J_2$-$J_3$ effective Heisenberg model,

$$H = J_1 \sum_{\langle i,j \rangle} \vec{S}_i \cdot \vec{S}_j + J_2 \sum_{\langle\langle i,j \rangle\rangle} \vec{S}_i \cdot \vec{S}_j + J_3 \sum_{\langle\langle\langle i,j \rangle\rangle\rangle} \vec{S}_i \cdot \vec{S}_j$$

where $J_1$, $J_2$, and $J_3$ denote the respective couplings between the NN, NNN, and third-nearest-neighbor Fe spins, and $S$ is the local magnetic moment on the Fe atoms. Considering the relatively small strain strength (~3%), we assume that the couplings $J$ remain nearly isotropic. By using the energy mapping method for the four AFM configurations (Néel, bicollinear, stripe, and dimer AFM in **Fig. S6**), we obtained the corresponding values of $J_1$, $J_2$, and $J_3$ as displayed in **Table I**. The values of $J_1$, $J_2$, and $J_3$ all increase under uniaxial compressive strain, with $J_3$ showing the most significant enlargement of approximately 16.8%. This may originate from the fact that under uniaxial compressive strain, the electronic density of states (DOS) at the Fermi level increases (**Fig. S8**), which can enhance the itinerancy of system and strengthen the exchange coupling between distant Fe spins. Actually, the energy difference between the trimer and dimer AFM states from the effective Heisenberg model is $(J_1 - 2J_2 + 2J_3)/3$ [17]. These results confirm that uniaxial compressive strain along the Fe-Fe NN direction could significantly tune the exchange coupling $J_3$ between the third-nearest-neighboring Fe atoms, thus stabilizing the dimer AFM order.

**Table I**. Exchange coupling constants $J_1$, $J_2$ and $J_3$ (in units of meV/$S^2$) between the Fe spins for ML FeTe without and with 3% (5%) uniaxial compressive strain applied along the nearest-neighbor (next nearest-neighbor) Fe-Fe direction. The values are obtained by calculating the energy differences among the stripe AFM, Néel AFM, dimer AFM and bicollinear AFM states.

| **Compressing direction** | **Strain (%)** | $J_1$ | $J_2$ | $J_3$ |
|---|---|---|---|---|
| NN | 0 | 53.4 | 38.0 | 18.5 |
| | −3 | 54.3 | 39.1 | 21.6 |
| | $\triangle J$ (%) | 1.7 | 2.9 | 16.8 |
| NNN | 0 | 53.4 | 38.0 | 18.5 |
| | −5 | 13.7 | 18.3 | 16.9 |
| | $\triangle J$ (%) | −74.3 | −51.8 | −8.6 |

Our findings demonstrate that the application of uniaxial strain along the Fe-Fe NN direction acts as a symmetry-breaking field that introduces a significant energy gap between these competing phases. By modifying the Fe-Fe bond lengths anisotropically, the strain differentiates the magnetic exchange couplings, effectively "freezing" the magnetic dimers into a coherent $\sqrt{2} \times \sqrt{2}$ lattice. This frustration-to-order transition is facilitated by the suppression of quantum fluctuations through lattice distortion. Furthermore, the selective stabilization of the bicollinear order under *ab*-axis compression versus the dimer order under *a'b'*-axis compression highlights a strong *J*-selective response. This confirms that the magnetic architecture of FeTe is not merely a byproduct of general lattice distortion but is governed by the precise symmetry and directionality of the strain tensor.

The dimer AFM can be hardly observed in thicker FeTe films that host superconductivity [11,12]. The larger magnitude and nonuniformity of the strain in few-FeTe films lead to more AFM domains than in the thicker films with similar stoichiometry. The local strain-induced AFM might hinder the superconductivity. This explains the increase of the superconducting transition temperature and sharpness with increasing film thickness.

In conclusion, this work discovers the tight interconnections between the emergent magnetic orders and local lattice distortions in few-layer FeTe films through a combination of high-resolution STM strain analysis and DFT calculations. Theoretically predicted dimer

AFM state, characterized by a $\sqrt{2}\times\sqrt{2}$ charge modulation on the top Te layer, is successfully stabilized by anisotropic strain along Fe-Fe NN directions. In contrast, the BiAFM order is exclusively confined by the uniaxial compression along Fe-Fe NNN directions. These findings establish a microscopic understanding of strain-engineered magnetic landscapes in iron-based chalcogenides and may have potential applications in spintronics devices.

## Methods

**Sample preparation.** FeTe films were prepared in an MBE system. The base pressure of the chamber was ~$1 \times 10^{-10}$ Torr. 0.5 wt.% Nb:STO substrates were annealed successively at 1050 °C for 40 minutes and 1150 °C for 20 minutes to obtain atomically flat surfaces. FeTe films were grown by coevaporating high-purity Fe and Te from standard Knudsen cells while the substrate was held at 270°C. FeSe buffer layers were grown by codepositing Fe and Se on the substrate held at 400 °C followed by annealing at 480 °C. All FeTe films were finally annealed within Te vapor to minimize interstitial Fe density[11].

**STM measurements.** *In-situ* STM measurements were conducted on a SI-STM-4K instrument (CIENSS Co., Ltd.). The sample temperature was 4.6 K with liquid helium cooling unless otherwise noted. A polycrystalline PtIr tip was used and calibrated on Ag islands before STM experiments. Images were scanned in the constant current mode. $dI/dV$ maps were acquired by scanning the image while adding oscillation on the bias voltage and extracting the demodulation from the tunneling current by a standard lock-in technique. The typical oscillation frequency is 973.2 Hz and the amplitude is 5 mV.

**First-principles calculations.** The electronic structures and magnetic properties of FeTe monolayer under uniaxial strain were studied by using the DFT calculations as implemented in the VASP package [18-20]. The fully spin-polarized electronic structure calculations were carried out by using the projector augmented wave (PAW) method [21]. The generalized gradient approximation (GGA) of Perdew-Burke-Ernzerhof (PBE) type [22] for the exchange-correlation potentials was adopted. The kinetic energy cut-off of plane wave basis was set to 520 eV. The convergence criterion for the forces on all atoms was set to 0.01 eV/Å. The STM simulations were performed to investigate the surface electronic structure of FeTe in both the bicollinear AFM and dimer AFM states. The VASPKIT package [23] was used to generate simulated STM images based on DFT calculations. We constructed a series of structures (**Figs. S5 and S6**) to investigate the effect of uniaxial strain applied along the nearest-neighbor and next-nearest-neighbor Fe–Fe bond directions on the magnetic states.

The lattice constant of pristine monolayer FeTe was set to 3.905 Å taking into account the tensile strain induced by the STO substrate.

## Data availability

All data supporting the findings of this study are available within the article and its Supplementary Information. Any additional requests for information on this study are available from the corresponding authors upon request.

## Acknowledgements

This work was supported by Quantum Science and Technology-National Science and Technology Major Project (2023ZD0300500), National Natural Science Foundation of China (Grant Nos. 12174443, 12434009 and 12304189), Beijing Natural Science Foundation (Grant Nos. 1242037, 1232035), Beijing Municipal Science & Technology Commission (Grant No. Z221100002722013). The theoretical work was also supported by the National Key R&D Program of China (Grants No. 2022YFA1403103 and No. 2024YFA1408601). Computational resources have been provided by the Physical Laboratory of High Performance Computing at Renmin University of China and the Beijing Super Cloud Computing Center.

## Author contributions

C.L. conceived and designed the research. H.X. prepared the samples. H.X., X.G. and C.L. conducted the STM measurements. J.J., K.L. and Z.-Y.L. carried out first principles calculations. H.X. and C.L. visualized and analyzed the data. H.X., X.G., H.L. and C.L. maintained the equipment. K.C. and C.L. supervised the project. H.L., K.L., Z.-Y.L. and K.C.

discussed on the results. C.L. wrote the manuscript with the input from all authors.

## Competing interests.

The authors declare no competing interests.

## Additional information

Supplementary information is available for this paper.

**Correspondence** and requests for materials should be addressed to Kai Liu, Kai Chang or Chong Liu.

Supporting Information

# Selective stabilization of antiferromagnetic orders in FeTe films via local strain engineering

Hao Xu[1,4,5,*], Jing Jiang[2,3,*], Xuesong Gai[1,4,5], Haicheng Lin[1], Kai Liu[2,3]✉, Zhong-Yi Lu[2,3], Kai Chang[1]✉ & Chong Liu[1]✉

[1] *Beijing Key Laboratory of Fault-Tolerant Quantum Computing, Beijing Academy of Quantum Information Sciences, Beijing 100193, China*

[2] *School of Physics, Renmin University of China, Beijing 100872, China*

[3] *Key Laboratory of Quantum State Construction and Manipulation (Ministry of Education), Renmin University of China, Beijing 100872, China*

[4] *Beijing National Laboratory for Condensed Matter Physics, Institute of Physics, Chinese Academy of Sciences, Beijing 100190, China*

[5] *University of Chinese Academy of Sciences, Beijing 100049, China*

*These authors contributed equally

✉e-mail: kliu@ruc.edu.cn; changkai@baqis.ac.cn; liuchong@baqis.ac.cn

## Note I. Rotational transformation for a second-order tensor

Let $\mathbf{T}$ be a second-order tensor. When the coordinate system is rotated by an orthogonal transformation matrix $\mathbf{R}$ (where $\mathbf{R}^{-1} = \mathbf{R}^{\mathrm{T}}$), the components of the tensor in the new (primed) coordinate system are given by

$$\mathbf{T}' = \mathbf{R}\,\mathbf{T}\,\mathbf{R}^{\mathrm{T}}$$

In two dimensions, the rotation matrix by an angle $\theta$ is

$$\mathbf{R} = \begin{bmatrix} \cos\theta & -\sin\theta \\ \sin\,\theta & \cos\theta \end{bmatrix}$$

Substituting this into $\mathbf{T}' = \mathbf{R}\,\mathbf{T}\,\mathbf{R}^{T}$ yields the explicit component transformations:

$$\begin{aligned} T'_{xx} &= \frac{T_{xx} + T_{yy}}{2} + \frac{T_{xx} - T_{yy}}{2}\cos 2\theta - T_{xy}\sin 2\theta \\ T'_{yy} &= \frac{T_{xx} + T_{yy}}{2} - \frac{T_{xx} - T_{yy}}{2}\cos 2\theta + T_{xy}\sin 2\theta \\ T'_{xy} &= \frac{T_{xx} - T_{yy}}{2}\sin 2\theta + T_{xy}\cos 2\theta \end{aligned}$$

Then,

$$T'_{yy} - T'_{xx} = \left(T_{yy} - T_{xx}\right)\cos 2\theta + 2T_{xy}\sin 2\theta$$

For $\theta = \pi/4$,

$$T'_{yy} - T'_{xx} = 2T_{xy}$$

This proves that twice the shear strain component in the *ab* coordinate is equivalent to the anisotropic normal strain in the *a'b'* coordinate. We also demonstrate it in Fig. S3 using the experimental data from Fig. 3.

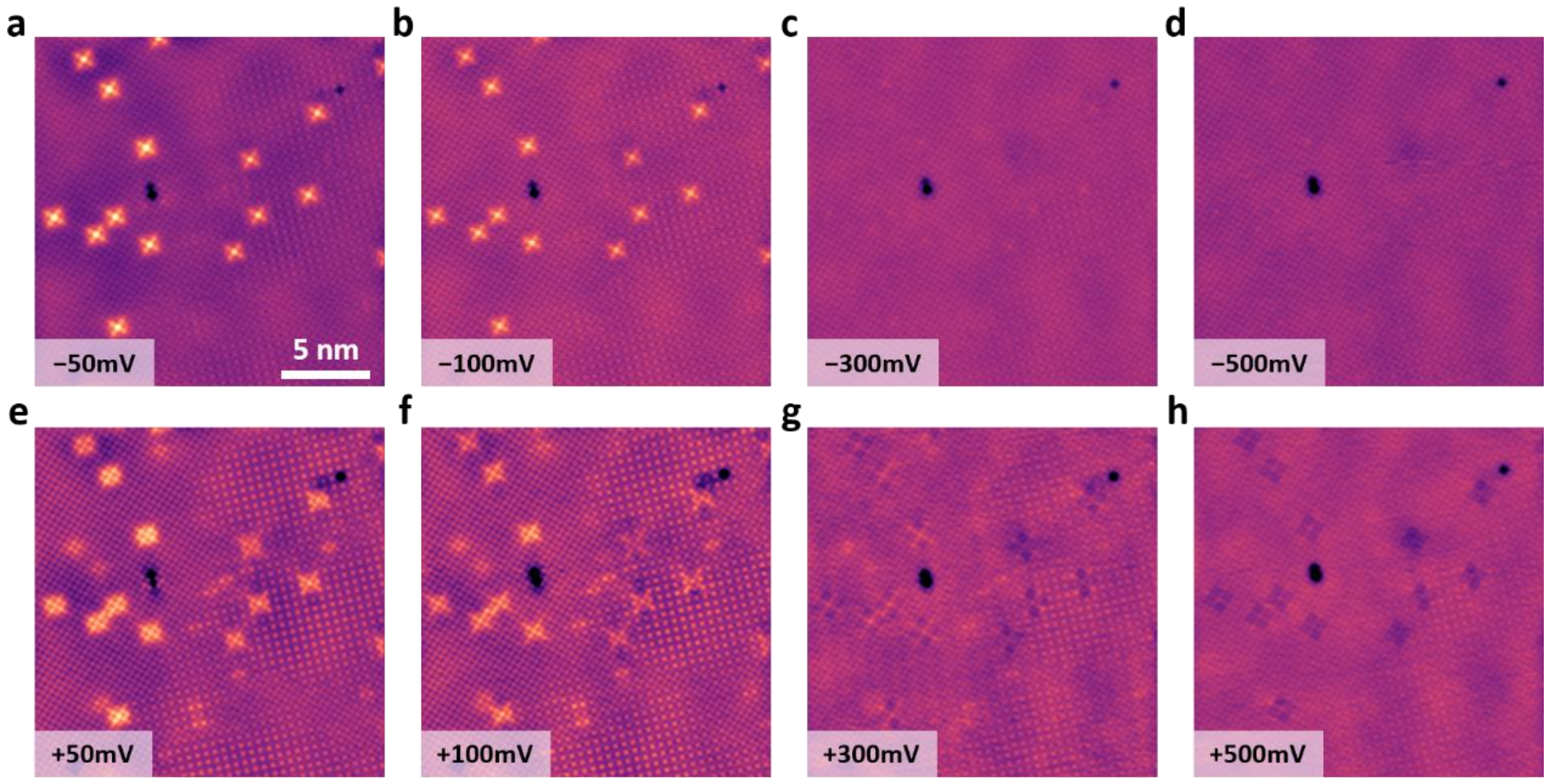

Figure S1. Bias-dependent topographic images taken at an area with dimer AFM phase. A polynomial background was subtracted. The $\sqrt{2} \times \sqrt{2}$ order gets weaker with increasing bias voltage.

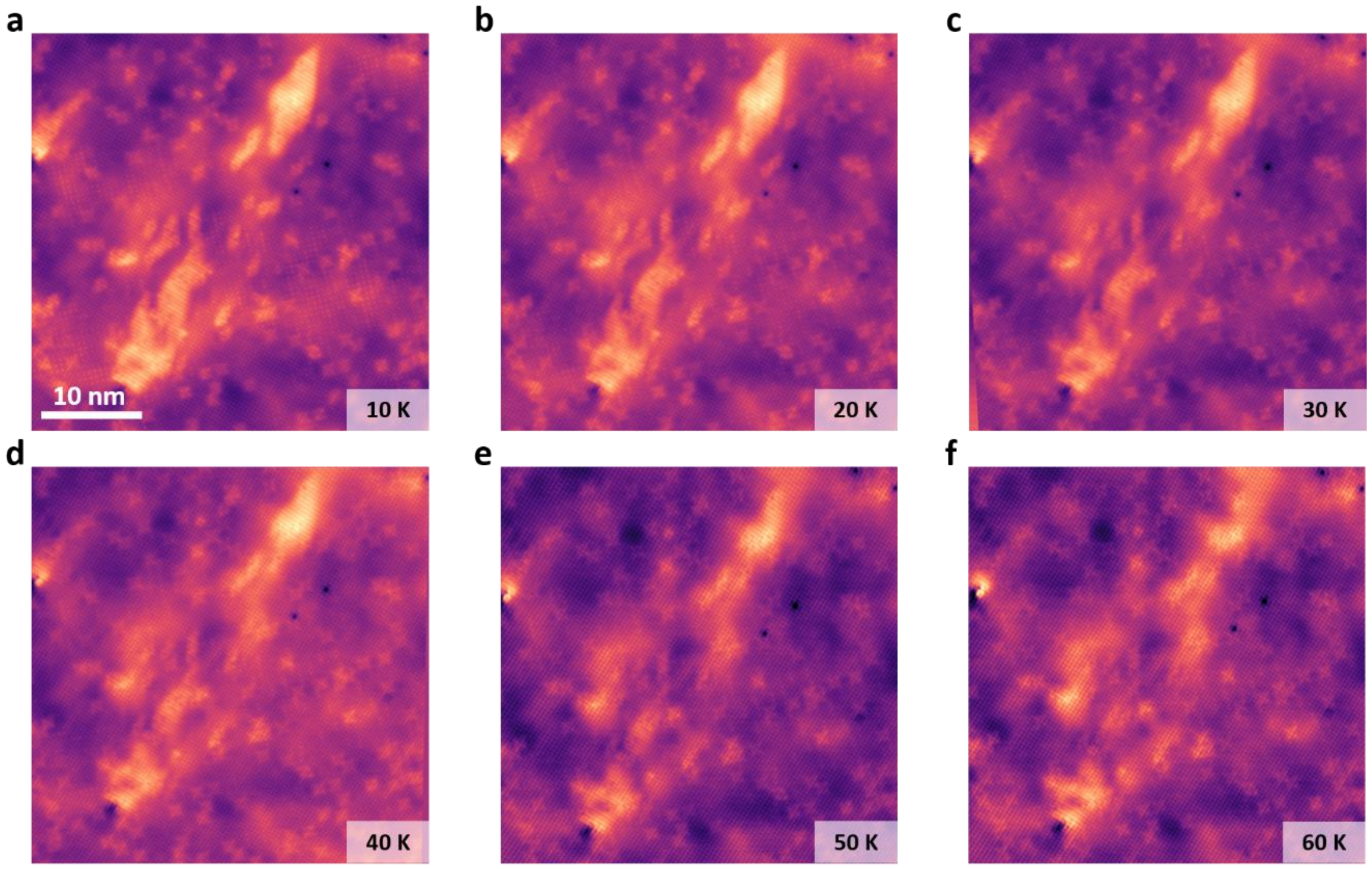

Figure S2. Temperature-dependent STM image taken at the same area as Fig. 1a ($V_s$ = 50 mV, $I_t$ = 200 pA).

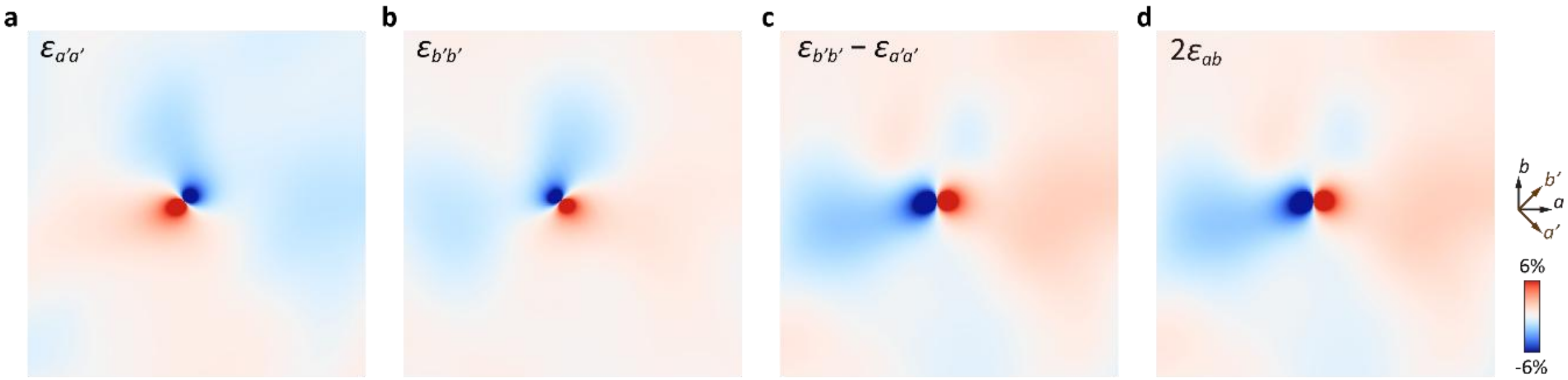

Figure S3. (a–b) Normal strain tensor components ($\varepsilon_{a'a'}$ and $\varepsilon_{b'b'}$) for Fig. 3b, extracted via geometric phase analysis in the $a'b'$ coordinate. (c) The anisotropic strain, $\varepsilon_{b'b'} - \varepsilon_{a'a'}$. (d) Two times shear strain component in the $ab$ coordinate, which is identical to (c).

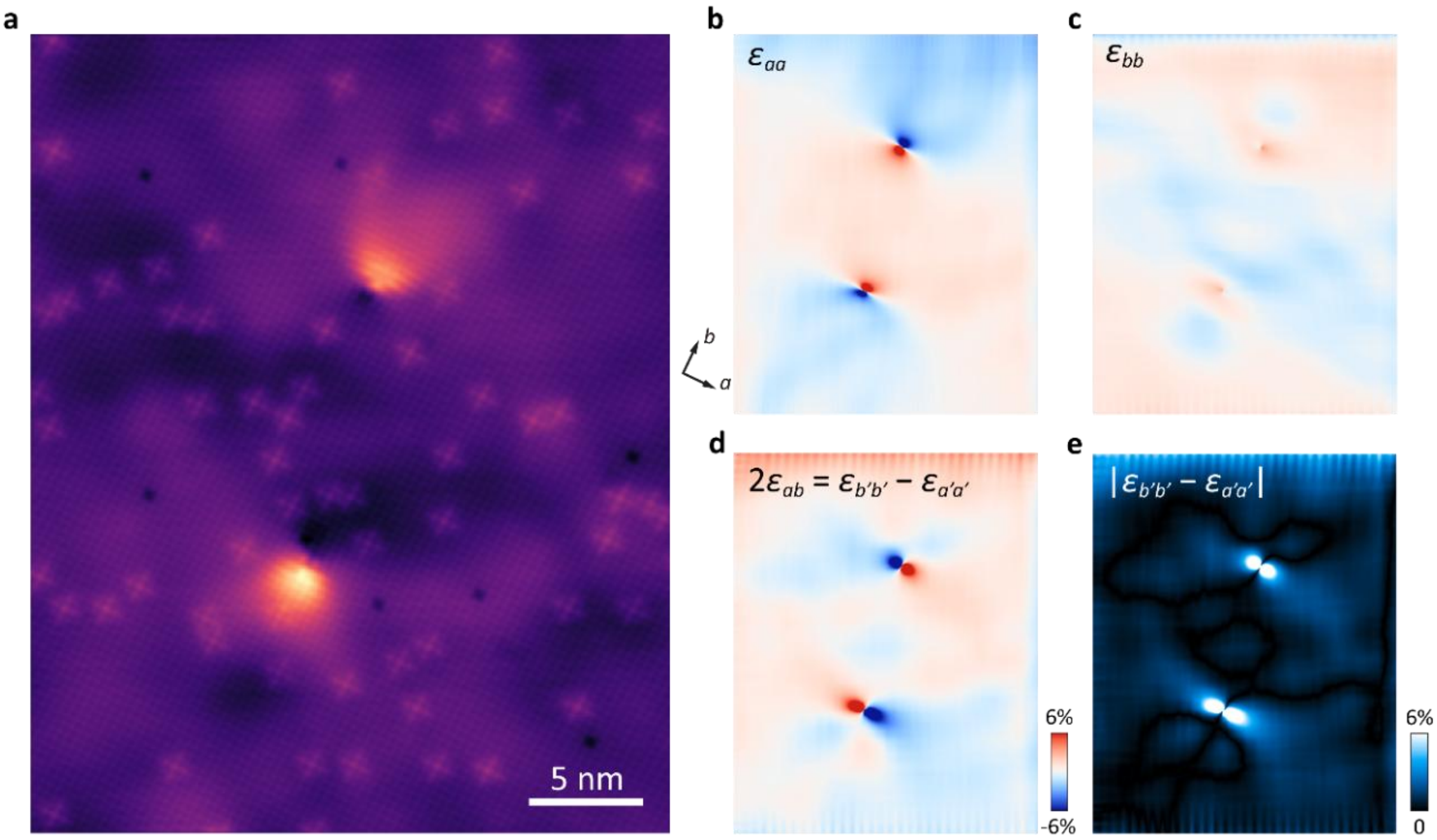

Figure S4. Strain field around the edge dislocations at 77 K. (a) STM image containing two edge dislocations with opposite directions ($V_s = -50$ mV, $I_t = 500$ pA). (b-e) Strain analysis similar to Fig. 3c-f.

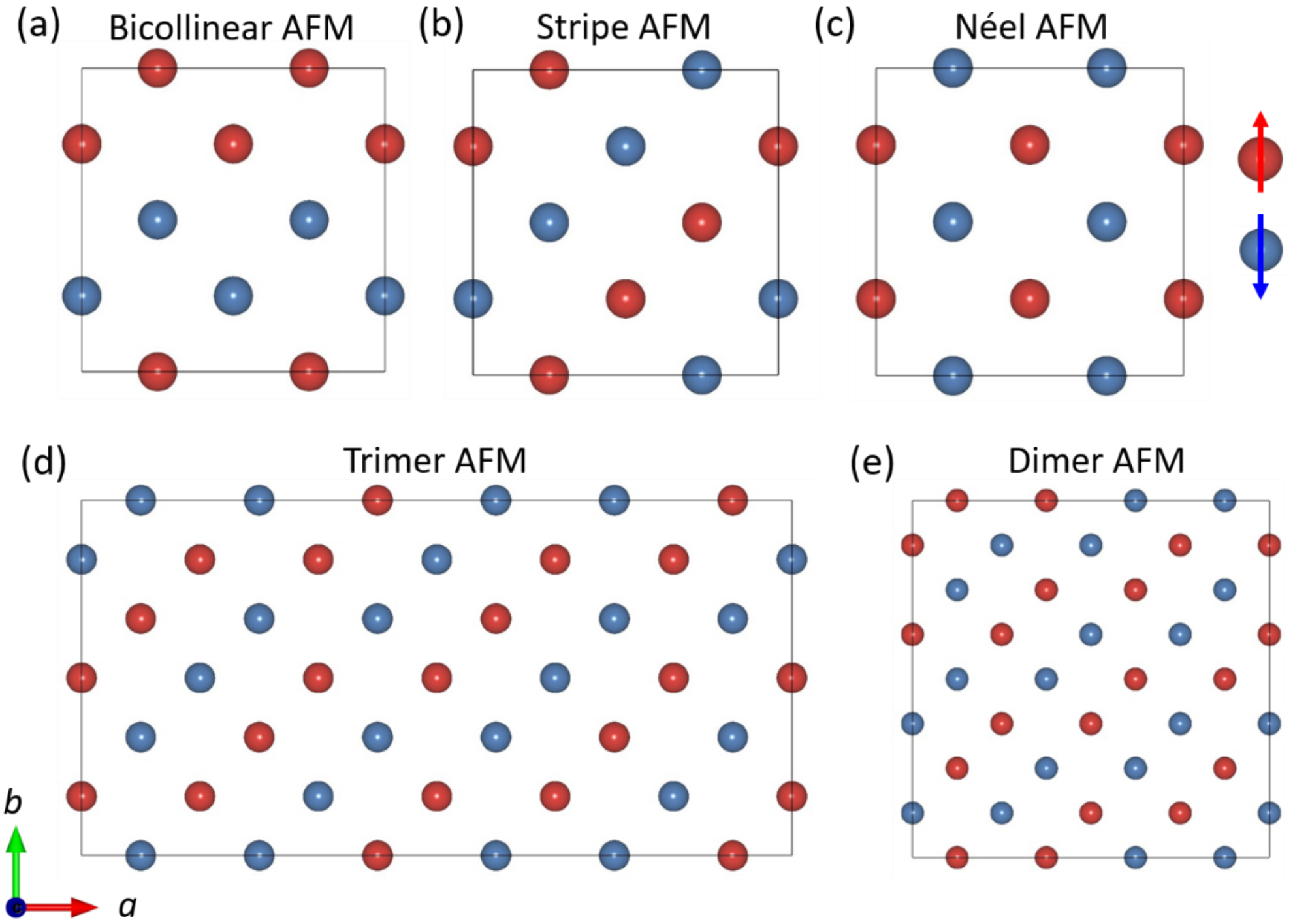

Figure S5. Five typical AFM states of FeTe under uniaxial strain along the next-nearest-neighbor Fe–Fe bond direction: (a) Bicollinear AFM, (b) Stripe AFM, (c) Néel AFM, (d) Trimer AFM, and (e) Dimer AFM states. The red and blue spheres represent Fe atoms with spin-up and spin-down orientations, respectively. Note that for the bicollinear AFM configuration, the uniaxial strain applied along the *a* and *b* directions are not equivalent. We define "bicollinear-a" ("bicollinear-b") for the bicollinear AFM state as the strains are applied along the *a* (*b*) directions.

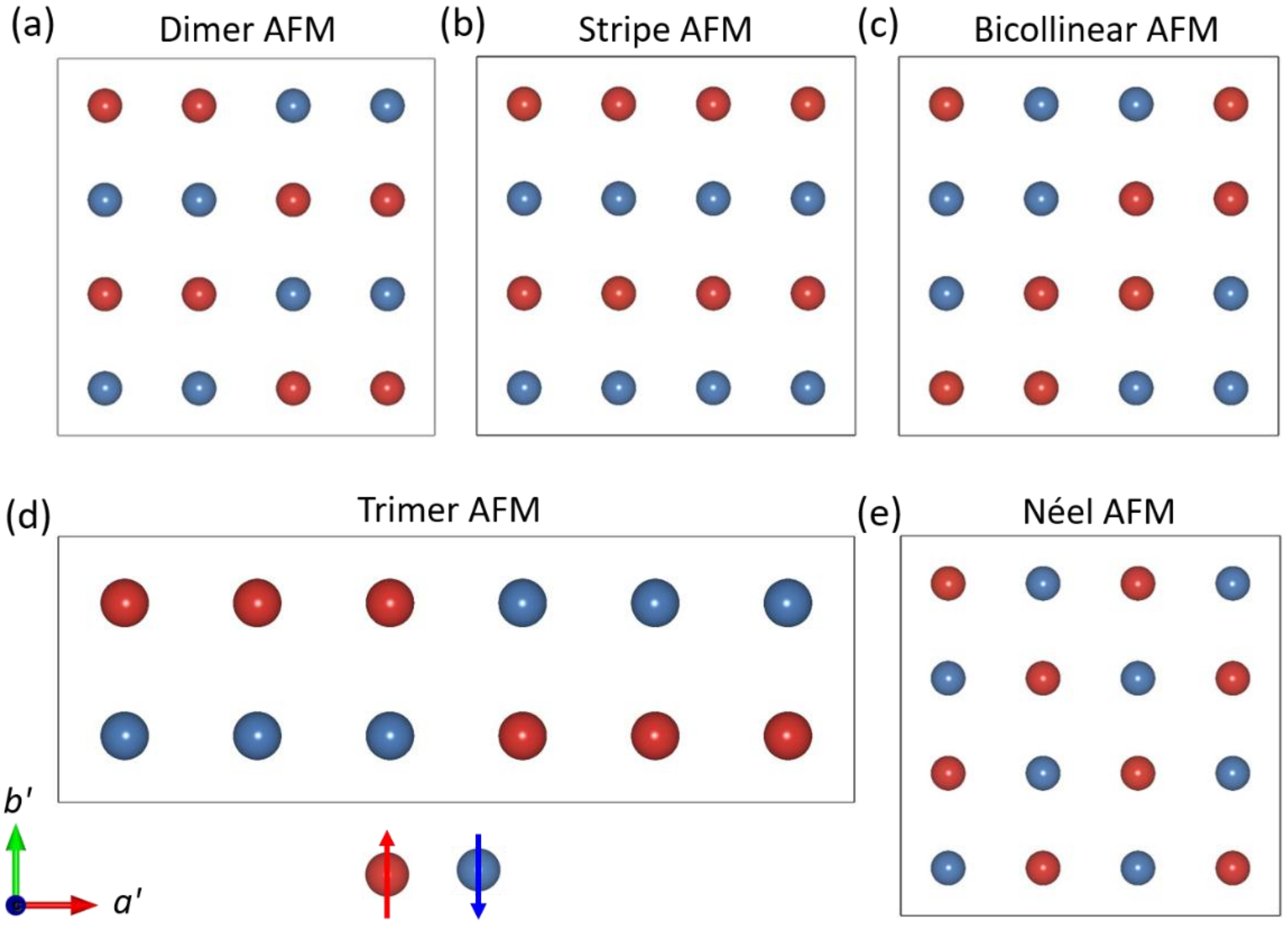

Figure S6. Five typical antiferromagnetic (AFM) states of FeTe under uniaxial strain along the nearest-neighbor Fe–Fe bond direction: (a) Dimer AFM, (b) Stripe AFM, (c) Bicollinear AFM, (d) Trimer AFM, and (e) Néel AFM states. The red and blue spheres represent Fe atoms with spin-up and spin-down orientations, respectively. Note that for the stripe, dimer, and trimer AFM states, uniaxial strains applied along the *a′* and *b′* directions are not equivalent. Accordingly, we define "dimer-a′" ("dimer-b′") for the dimer AFM state as the strains are applied along the *a* (*b*) directions.

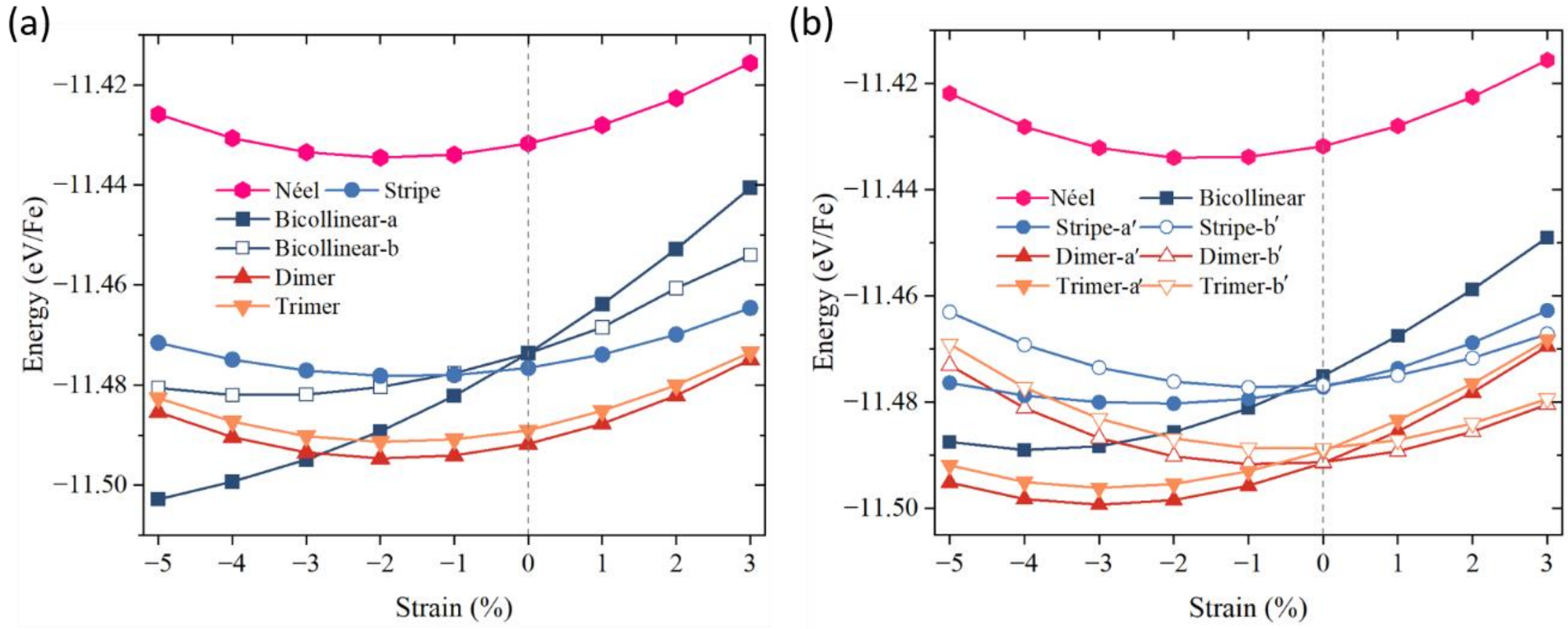

Figure S7. The comparison of total energies for different magnetic states of monolayer FeTe under uniaxial strain applied along (a) the nearest-neighbor Fe–Fe bond direction and (b) the next-nearest-neighbor Fe–Fe bond direction. The -*a*, -*b*, -*a′* and -*b′* denote the directions along which the uniaxial strains are applied.

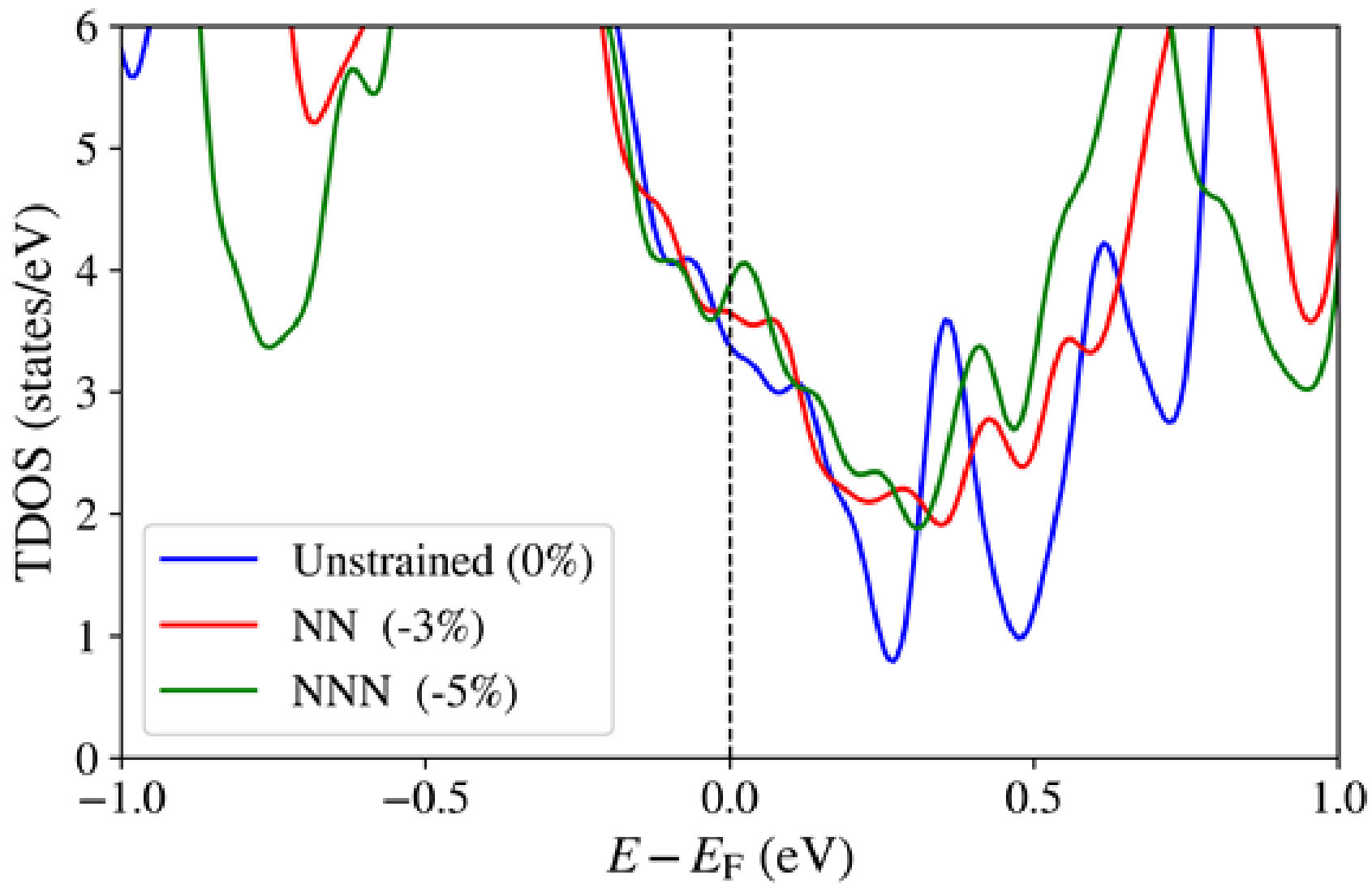

Figure S8. Calculated total density of states (TDOS) of ML FeTe without strain and under NN (−3%) and NNN (−5%) uniaxial compressive strain in the nonmagnetic states.